\pdfoutput=1
\documentclass{vgtc}                          

\ifpdf
  \pdfoutput=1\relax                   
  \usepackage{graphicx}                
  \DeclareGraphicsExtensions{.pdf,.png,.jpg,.jpeg} 
\else
  \usepackage{graphicx}                
  \DeclareGraphicsExtensions{.eps}     
\fi%

\graphicspath{{figures/}{pictures/}{images/}{./}} 

\usepackage{microtype}                 
\PassOptionsToPackage{warn}{textcomp}  
\usepackage{textcomp}                  
\usepackage{mathptmx}                  
\usepackage{times}                     
\usepackage{cite}                      
\usepackage{tabu}                      
\usepackage{booktabs}                  

\onlineid{0}

\vgtccategory{Research}

\vgtcinsertpkg

\title{Fixation and Creativity in Data Visualization Design: Experiences and Perspectives of Practitioners}

\author{Paul Parsons\thanks{e-mail: parsonsp@purdue.edu}
\and Prakash Shukla\thanks{email: shukla37@purdue.edu}
\and Chorong Park\thanks{email: park1187@purdue.edu}}
\affiliation{\scriptsize Purdue University}

\abstract{Data visualization design often requires creativity, and research is needed to understand its nature and means for promoting it. The current visualization literature on creativity is not well developed, especially with respect to the experiences of professional data visualization designers. We conducted semi-structured interviews with 15 data visualization practitioners, focusing on a specific aspect of creativity known as \textit{design fixation}. Fixation occurs when designers adhere blindly or prematurely to a set of ideas that limit creative outcomes. We present practitioners' experiences and perspectives from their own design practice, specifically focusing on their views of (i) the nature of fixation, (ii) factors encouraging fixation, and (iii) factors discouraging fixation. We identify opportunities for future research related to chart recommendations, inspiration, and perspective shifts in data visualization design.   %
} 

\CCScatlist{
  \CCScatTwelve{Human-centered computing}{Visu\-al\-iza\-tion}{}{};
}

\begin{document}


\firstsection{Introduction}

\maketitle
Data visualization designers often need to produce solutions that are both novel and useful. This combination of attributes forms the standard definition of creativity \cite{runco2012standard}, and is commonly used to define creativity in design \cite{Crilly_2019,sarkar2011assessing}. Design creativity has been studied in numerous design disciplines, including interaction design \cite{Frich_MacDonald_Vermeulen_Remy_Biskjaer_Dalsgaard_2019}, instructional design \cite{Clinton_Hokanson_2012}, product design \cite{Yilmaz_2011}, industrial design \cite{Dorst_Cross_2001}, and architecture \cite{goldschmidt2005good}. Researchers in these fields have investigated many aspects of creativity, including, among others, the role of constraints \cite{onarheim2012creativity}, the nature of a `good' idea \cite{goldschmidt2005good}, where creativity is located \cite{gero2011future}, the influence of environmental information \cite{COLLADORUIZ2010479}, and numerous cognitive aspects relating to creativity \cite{kim2007underlying,gero2015studying,casakin2011cognitive,bonnardel2000towards}. 

Although the visualization community has not studied creativity in the same breadth or depth as other design disciplines, there have been some initial contributions. With respect to the visualization design process, Vande Moere and Purchase argue for a view that includes both creative and engineering aspects \cite{Moere_Purchase_2011}. In proposing the Design Activity Framework, McKenna et al. \cite{McKenna_Mazur_Agutter_Meyer_2014} acknowledge the role of creativity in the visualization design process, and argue that previous design process models have lacked this emphasis. Regarding the promotion of creativity, Goodwin et al. \cite{Goodwin_Dykes_Jones_Dillingham_Dove_Duffy_Kachkaev_Slingsby_Wood_2013} have investigated several techniques, such as analogical reasoning and constraint removal. Kerzner et al. \cite{Kerzner_Goodwin_Dykes_Jones_Meyer_2019} have proposed a framework for ``creative visualization-opportunities workshops'' inspired by creative problem solving workshops. Mendez et al. \cite{Mendez_Hinrichs_Nacenta_2017} investigated the role of top-down and bottom-up approaches to visualization construction, arguing that bottom-up approaches can promote creativity. Others have developed techniques that may promote creativity, such as using cards \cite{he2016v}, sketching and ideation activities \cite{Bressa_Wannamaker_Korsgaard_Willett_Vermeulen_2019}, and worksheets \cite{Roberts_Headleand_Ritsos_2016,mckenna2017worksheets}. While these studies provide valuable contributions to promoting creativity in data visualization, there are still many topics that need to be explored, especially relating to the nature of creativity and factors that may hinder or promote it.

One factor influencing creativity that has been studied across multiple disciplines is \textit{design fixation}. Fixation occurs when designers adhere blindly or prematurely to a set of ideas---concepts, features, representations, solutions---that limit creativity in the design process \cite{Jansson_Smith_1991}. Studies across design disciplines have shown that fixation exists and impacts design cognition in significant ways, yet it is not well understood for data visualization. One of the only papers to explicitly discuss fixation in data visualization is from Eggermont et al. \cite{Eggermont_Knudsen_Pusch_Carpendale}, in which they present a functional co-occurrence visualization tool with the aim of avoiding design fixation. However, their short poster abstract does not appear to have been extended.

In this paper we present findings from semi-structured interviews with 15 data visualization practitioners, in which we asked them about their experiences and perspectives on fixation, including what might encourage or discourage it. We identify opportunities for future research related to chart recommendations, inspiration, and perspective shifts in data visualization design.


\section{Design Fixation}
\label{sec:designfixation}
Fixation has been studied extensively by psychologists, often in reference to problem solving and insight \cite{smith_incubation_1991,bowden2005new,duncker1945problem}. Inspired by this work, Jansson and Smith\cite{Jansson_Smith_1991} defined \textit{design fixation} as `blind adherence to a set of ideas or concepts limiting the outputs of conceptual design'. They performed a series of experiments in which participants were given a design problem, and one group was shown an existing solution (fixation group) while the other was not (control group). Results showed that the fixation group generated solutions with less variation than the control group, indicating the existence of fixation in the design process.

Since this early work, there have been studies on design fixation across a wide variety of disciplines, including engineering design \cite{linsey_study_2010,viswanathan_study_2014,toh_mitigating_2014}, architecture \cite{Goldschmidt_variances_2006}, industrial design \cite{cheng_new_2014}, product design \cite{hwang_mitigating_2020}, and service design  \cite{moreno_overcoming_2016}. Researchers have employed various methods to study design fixation, including think aloud protocols of sketching activity \cite{goldschmidt_dialectics_1991}, showing partial photographs to designers \cite{cheng_new_2014}, and using physical and non-physical environments to investigate effects on fixation during prototyping \cite{Youmans_effects_2011}. 

Though there have been many empirical lab- and classroom-based investigations of design fixation, there has not been as much research using qualitative approaches focused on real-world design settings. Crilly \cite{Crilly_2015, crilly_methodological_2019} has advocated for more methodological pluralism in the study of fixation, including qualitative approaches and observational field studies with practitioners. Although a focus on practitioner perspectives in general is growing in the visualization literature (e.g., \cite{Parsons_Gray_Baigelenov_Carr_2020,Parsons_Shukla_2020,bigelow_reflections_2014,Hoffswell_Li_Liu_2020,Mendez_Hinrichs_Nacenta_2017,walny_data_2020,Alspaugh_Zokaei_Liu_Jin_Hearst_2019,Parsons_2021,brehmer2021jam}), it has not historically been a strong area of focus. Our work provides a practice-led, qualitative approach to understanding fixation in a data visualization context.




\section{Method}
We interviewed 15 data visualization practitioners about their experiences and perspectives on fixation. Recruiting was done via social media, the DataVis Society’s Slack workspace, the InfoVis email list, and our personal networks. We also contacted more than 200 individuals and more than 30 agencies. This discussion on fixation happened at the end of a larger interview aimed at understanding participants' design process (see \cite{Parsons_2021} for more detail). Participants were given the option to continue and discuss their views on creativity, to which 15 of the original 20 participants agreed. The discussion of fixation took approximately 15 minutes. Table \ref{tab:participants} provides self-reported characteristics of the participants. 

Interviews were semi-structured and conducted remotely via videoconferencing. Participants were first asked if they could describe a project in which they were creatively stuck and then made a breakthrough. The interviewer then introduced the concept of fixation. In line with Crilly et al. \cite{Crilly_2015}, this included a high-level description of Jansson and Smith’s \cite{Jansson_Smith_1991} experiments with the bike rack problem (see Section \ref{sec:designfixation}), which served as an example of design fixation and its effects. Participants were asked about their thoughts on what might encourage fixation and what might discourage fixation in the context of data visualization design.

Transcripts were analyzed using a hybrid thematic analysis approach \cite{fereday2006demonstrating}, incorporating both data-driven inductive coding and a top-down a priori approach. Our top-down approach was guided by previous literature on design fixation \cite{Crilly_2015}. We specifically looked for data on (1) descriptions of fixation in participants' own words; (2) factors that may encourage or lead to fixation; and (3) factors that may discourage or help overcome fixation. The three researchers went through several rounds of bottom-up coding independently with these high-level categories in mind. We met regularly to discuss the codes, eventually coming into alignment on the final codes. 

\begin{table}
\begin{tabular}{ p{0.05\columnwidth} | p{0.42\columnwidth} | p{0.09\columnwidth} | p{0.13\columnwidth} | p{0.06\columnwidth} }
\textbf{ID} & \textbf{Job Title} & \textbf{Exp. (yrs)} & \textbf{Highest Degree} & \textbf{G} \\ 
\hline
P1 & DataVis Designer & 5-7 & B & M \\
P2 & DataVis Designer & 5-7 & B & F \\
P3 & DataVis Designer & 8-10 & B & M \\
P4 & Sr. DataVis Dev & $>$10 & D & F \\
P5 & Data Communicator & 2-4 & M & M \\
P6 & DataVis Designer & $>$10 & M & M \\
P7 & DataVis Designer & 5-7 & M & F \\
P8 & DataVis/UX Designer & 2-4 & M & F \\
P9 & Graphics Editor & 8-10 & B & F \\
P10 & Sr. UX Design Lead & 8-10 & D & M \\
P11 & DataVis/UX Designer & $>$10 & M & M \\
P12 & Data Architect & $>$10 & M & M \\
P13 & Sr. UX/DataVis Designer & 5-7 & M & F \\
P14 & DataVis Designer & 5-7 & M & F \\
P15 & DataVis Journalist/Designer & 5-7 & M & M \\

\hline
\end{tabular}
    \caption{Self-reported characteristics of our participants: job title, years of experience, highest degree (Bachelor's, Master's, or Doctoral degree), and gender. ~\label{tab:participants}}
\end{table}

\section{Findings}

\subsection{Nature of Fixation}


All participants reported experiencing fixation in their professional work and were able to describe a project they had worked on in which fixation was an issue. Many participants described fixation as being a common issue in their own practice. In describing fixation, participants made references to getting `stuck', `hitting a wall', having difficulty `letting go', having a `blockage', and hitting a `dead end'. Some participants described the feeling of being fixated. P10, describing an instance of fixation, stated ``\textit{I certainly felt like I was throwing myself at a brick wall over and over and over again until I had some notion of how to get over that wall.}'' P1 described the feeling of fixation as needing to ``\textit{get my head out of that focused, that super-focused state, where I can't consider other things.}'' P7 described the attachment people feel towards their design ideas, noting the solution as ``\textit{you have to kill your darlings.}''

Some participants described fixation as getting stuck on an initial idea. For example, when asked if they experience fixation, P6 replied ``\textit{No question about it---and I've experienced in two different ways. One is I come up with an idea like immediately during the phone call with the client, and then I'm sort of fixated on that initial idea and can't veer away from it or just sort of tend not to. And I've also had the experience [\ldots] where maybe the client says, `Oh, we were thinking of doing X, Y, Z'---and now I'm stuck with their idea.}'' 

Other participants interpreted fixation more broadly. P14 described fixation as not being able to find a better solution than one they currently have: ``\textit{I have cases where I never find a good solution. I find that I'm never totally happy with it and I just stick with the least bad version.}'' P15 described an instance of fixation as being about ``\textit{the story and the structure of the story I wanted to tell''}, going on to describe the effect of re-framing the problem ``\textit{finally I found an angle [\ldots] the framing, that was really the angle I needed to build a story. And that was kind of a breakthrough. Before that I was really a bit lost until I found that angle.}'' 



\subsection{Factors Encouraging Fixation}

\textbf{Chart Recommendations} Multiple participants noted the existence of chart recommendations from software as promoting premature adherence to a design idea. For instance, P15 noted ``\textit{a lot of the people making visualizations are constrained by the chart types they have in their software [\ldots] by the all the chart types that Excel can offer, for example, and they don't know how to break out of this constraint of the software they're using.}'' Similarly, P7 stated ``\textit{ I think if you invest in a charting tool and it shows a visualization type, then you have like these borders to other visualization types because you need to completely start your visualization from scratch when you choose another visualization type.}'' P13 noted that ``\textit{[\ldots] like excel, these are the 10 types of charts that you can use. I think those can sometimes be bad because then people feel limited to them and you don't think about the 20 types of targets that excel doesn't do. And so anything that's sort of like limits your possibility of tools in general.[People say] Oh, I have to use Tableau I have to use, I can only design within this framework.}'' 

\textbf{Missing the Big Picture.} Participants mentioned being too focused on details and not the `big picture'. P12 described being ``\textit{so deeply involved in the details, and the pieces and parts that you---for whatever reason---have a hard time sitting back and looking at the whole thing in a more relaxed way. That will get people stuck.}''  P11 described a strategy he uses to avoid getting stuck in the details: ``\textit{Sometimes I've just taken the biggest marker I have and done designs with that. I mean the overly huge one that doesn't let you do any details, because if you are kind of stuck in the details, then you need to kind of somehow dig yourself out of it.}'' P14 mentioned getting stuck on details while programming: ``\textit{in the programming phase [\ldots] that's where it usually gets stuck on the details.}''


\textbf{Effort and Attachment} Participants described the amount of effort required for creativity in design as potentially encouraging fixation. P14 described this as being due to emotional investment in their work ``\textit {I'd have definitely have a lot more emotional difficulty to letting go of something that you've worked out for a while and are fixated on.''} P6 described experiencing the phenomenon of sunk cost, where ``\textit{You get invested in this thing you've invested time in and you just don't want to let go of it.}'' Related to the issue of chart recommendations above, P13 emphasized the ease of selecting from predetermined chart types ``\textit{Here are your suggested chart types okay, I can point and click and like, change it. And that's super easy.}''

\textbf{Existing Practices and Habits.} Some participants described the role of prior experience and existing practices or habits in influencing fixation in design. P4 described the influence of reusable components in adopting ideas too early ``\textit{We've got some reusable components, and that makes you kind of go to those first. Like, let's go to our library and see---so certainly I think that's something that, you know, I've experienced and our team has experienced.}'' P13 mentioned the role of having solved similar problems previously as leading to fixation: ``\textit{Because you think you've solved this problem before, you think it's the same problem.} P4 described experiencing fixation as a result of relying on prior success, ``\textit{ Everything we do is sort of building on what we've already done. You know, we would make one app and it goes out there, it's popular, so they're like, hey, make another, make one like that for this other survey because you've already got your original thing. It is very easy to get stuck on.}'' P1 described the role of rigid `best practices' in influencing the adoption of ideas as ``\textit{A lot of people get stuck on best practices and that being the stop of it, like this is the best practice [\ldots] I think best practices can hold people back if they see it as like they have to do this.}''


\textbf{Precedent.} Many participants described the role of existing visualizations in encouraging fixation. Participants were aware of their own propensity to be influenced by prior visualization designs and described strategies they had developed. P15 described their strategy as ``\textit{I don't want to look too closely at their [other visualization designers] stuff because I don't want to get stuck just building that.}'' P6 noted that looking at existing visualizations can encourage fixation: ``\textit {Yeah. You might get fixated on this cool thing on D3 and it's just not the best idea.''} P3 described a similar strategy as `` \textit{I made a conscious decision when I started a new project not to look at other people's work to get some inspiration. So right now, I'm not looking so much on a site anymore when I do projects because I did notice that you tend to come up with similar solutions.}''


\textbf{Client Influence.} Some participants described fixation arising from client influences, largely due to clients changing their mind, sharing their own ideas about the design, and brand guidelines that need to be applied. P6 described clients' early ideas as encouraging fixation, and a strategy of avoiding visual content from clients to prevent fixation: ``\textit {Oh, we were thinking of doing X, Y, Z, and now I'm stuck with their idea. And actually actively try to avoid that when clients have a visual content, I usually try to encourage them to not share it and let's not talk about that.''} P4 mentioned the role of client expectations as ``\textit {something that we hear a lot is like, well our clients are used to looking at it this way, and so that's certainly something that enforces that kind of fixation.}'' P6 also shared their experience of clients promoting adherence to an idea they've seen somewhere else: ``\textit{What might lead to it [fixation] in datavis specifically is---8 times out of 10---is charts, it's like, `oh, I think we should do a bar chart. I was thinking we should do a pie chart' [\ldots] or they saw this really cool thing in the New York times or wherever else and they say, `Oh, I saw this wonderful thing over here. Can we do that with this data?}''




\subsection{Factors Discouraging Fixation}

\textbf{Incubation.} Participants discussed how putting the problem aside for a while could help overcome fixation. P7 puts it as ``\textit {Taking a break, looking at other stuff. Even sometimes just taking, if I can, a couple days away, not thinking about it. And then somewhere in my subconscious something will click. And I will be able to like, oh, this was the problem. Let's try it this way instead.}". P2 highlighted the importance of diverting the mind from the problem and says the ``\textit {only thing I do when I know that I've gone down a path that's ineffective, or that I'm too fixated, is I just take a break---and that could be like I step away and do something for a few hours, or do a completely different project for weeks and come back with a fresh perspective.}'' Some participants talked about unconscious moments of insight, even in states of sleep, as P9 describes: ``\textit{I dunno, sometimes with things where I'm sleeping in the middle of the night, wake up and like, oh, I know exactly how to redo this.}'' 

\textbf{Experimenting and Creative Exercises.} Participants shared various strategies they employ to avoid or overcome fixation and promote creativity. Some participants described experimenting with different visual encodings as a creative exercise to overcome being fixated on one encoding. For instance, P3 described ``\textit{these kind of `what if' questions, like what if I would show it as line instead of circles or whatever. Because I think one of the characteristics of data visualization is that, especially because of this very iterative approach and exploratory approach at the small scale, you can try things and then you see things happen again. And sometimes, or often, these could be things that you didn't expect and these kinds of things can generate new ideas that you could try out.}'' Similarly, P1 described their strategy as ``\textit{pick two encoding and mash them together to see what sort of thing you can create from that.}"


Participants described other creative exercises, such as P4's  ``\textit {Coming up with 10 solutions to something [\ldots] eventually you start coming up with stuff that's weirder and that might lead off into something.''} P6 described sketching as a way to keep early ideas flexible and avoid fixation, stating that ``\textit {I try not to do it [look at existing visualizations] until after I got my initial sketches out.}'' P3 described a training exercise of asking people to create intentionally bad designs, reflecting that ``\textit{it's very interesting to see that once people start doing this exercise, they sometimes cannot stop working on it because they have so many ideas, and sometimes people are laughing at themselves because they come up with nice ideas. And so this is really one of the most fun exercises because people come up with the most crazy ideas.}''. P9 described an exercise of trying to do the opposite of an existing visualization for a similar problem that they were working on ``\textit{what if I showed the opposite of this? So instead of being like `what are the most words that the candidates used?' I'd be like, `what words didn't they use?'}'' P11 described a way to avoid early attachment to ideas as ``flipping the brief''---referring to a strategy for challenging the initial client request and adjust the framing of the problem.

\textbf{Inspiration.} Participants highlighted various ways they get inspired for their design work. One method was to look for inspiration in existing visualizations. P12 described a situation where a client wanted an `ugly' dashboard to become `visually appealing'. P12 was not sure what to do, so ``\textit{went into a sort of hysterical, `inspire me' mode}'' and stumbled across a waffle chart. P12 then described the result: ``\textit{so that [the waffle chart], combined with a better color palette than they had before, turned into pretty compelling results that they liked---and that got me out of the woods.}'' However, this is one of the trickiest strategies to follow with respect to fixation, as some participants indicated becoming fixated as a result of looking at existing work. Participants acknowledged this and discussed strategies for looking at existing work that was not too similar to the current problem they were working on. As P9 said ``\textit{I wouldn't look at any of their [a direct competitor] stuff because then I would feel like I was going to create something like what they were doing}.'' P3 noted that they will ``\textit{still look at other people's work, not related to a specific project that I'm working on.}'' P4 described a strategy of looking broadly at existing visualizations ``\textit{I'll go to a few different sites like Flowing Data or Xenographics that just have lots of examples of different visualizations, different chart types, and just browse that for half an hour. And that can help me kind of jog out of being stuck on something.}'' 

Some participants, especially P2 and P9, went beyond data visualization examples to look for inspiration in art and nature. P9 described ``\textit{going to the art museum and seeing a renaissance painting, or like this 1970s weird something. I went to this conspiracy UFO exhibit, but it's just like \ldots it [the solution] can be a shape---it could just be like a weird shape. I see. And I'm like, `oh like that's a connection I can make'.}'' P2 described getting inspiration as ``\textit{a lot of the inspiration is just kind of like the world around me---nature, other people's art. Especially, I like taking inspiration from art---not other people's datavis--but other people's art.}'' P2 described a project where they were trying to create a visual encoding using a flower metaphor, but were getting stuck on the idea: ``\textit{I just couldn't lay it out in a pleasing way. And then, I was in Tokyo at the time, and I stepped out and took a walk. And it was April, so there was a bunch of cherry blossoms and, as I was staring out at the cherry blossom, I'm like: `Oh yeah, flowers grow on trees'. And so if I just organize them as tree branches, it would look much more aesthetically pleasing---and then it just unblocked me immediately.}''

\textbf{External Input.} Many participants described the value of external input in overcoming fixation. P7 noted ``\textit{having a co-worker discussing this with me, and basically the question that they ask always gets me out of it [fixation]}.'' P6 spoke about seeking informal input from their spouse whenever being stuck creatively. P5 includes feedback as a part of their process and says, ``\textit{in my process, what helps me avoid it [fixation] is getting others’ input---that's input from users, input from coworkers---getting a sense of what other people are thinking and how they respond to something helps prevent me from getting locked into the idea that my idea is the only solution}."

\textbf{Changing Perspective.} Some practitioners advocated for deliberately thinking in a different way or changing their literal or mental perspective. P6 described their strategy of trying to think from a beginners' perspective: ``\textit{try to remember what it was like to not know anything, and try to walk into every client engagement as though you don't know anything about what they're doing or about even almost data visualization..}'' Some participants described a literal change of perspective, such as P12 suggesting to ``\textit{Take an image and look at it upside down to get a sense of it from a compositional perspective outside of the details of what it's actually a picture of [\ldots] one of the things you need to do is cock your head or go away and come back or look at it from across the room.}'' P2 speculated about the value of a technique that's often used in art education, where ``\textit{one of the things that you would do is flip your artwork upside down. You really give yourself a different perspective.}''


\section{Discussion}
Fixation was universally experienced by our participants. This is consistent with research on fixation in other design disciplines. Most participants could recognize fixation when it was described and could easily provide a personal example. Many, although not all, were aware of their own encounters with fixation and had developed conscious strategies to avoid or overcome it. Some participants could describe ways that they had overcome fixation, although they were not always deliberate strategies that had been developed. For instance, participants described serendipitous insights from nature, art galleries, existing data visualizations, and even dreams. These are consistent with prior research on fixation and creativity. Researchers have studied the role of both conscious and unconscious incubation, noting that the mind may still be making associations when people are attending to other things, including while sleeping \cite{cai2009rem,ritter2014creativity,zhong2008merits}.

Although fixation has been studied in other design disciplines, its nature and effects in data visualization are not known. Visualization practitioners need to engage in creative work that is not the same as the work that engineers or interaction designers may engage in. For instance, they need to deal with issues relating to data, visual encodings, human perception and cognition, and client and stakeholder needs---a unique set of issues not faced by designers in other disciplines \cite{walny_data_2020,Parsons_Shukla_2020}. It is known in general that looking at existing solutions can lead to fixation \cite{Jansson_Smith_1991}. However, we do not know exactly what that means for data visualization. What about visualizations based on partially or entirely different data types? Or visualizations that are static vs. interactive? What about the role of visualization grammars vs. chart templates? The answers to such questions cannot be easily transferred from other design disciplines, and necessitate their own studies within a visualization context. Our work does not attempt to answer these questions, but rather provides an initial exploration into this space, drawing connections to literature in other fields while also offering preliminary insights from a practice-led data visualization perspective.

Our findings surface questions about how to avoid or overcome fixation for data visualization designers. For instance, the role of chart templates and suggestions in software tools is a challenging topic. Participants noted that templates can be too easy to select and can promote adherence to a solution too early. P15 noted that templates can lead to fixation, while ``\textit{the grammar of graphics gives you a lot of flexibility and a lot of different ways of expressing your data [\ldots] it's definitely a framework that allows for creativity.}'' This is a topic that spans more than one of our factors that encourage fixation---at least \textit{chart recommendations}, \textit{effort and attachment}, and \textit{existing practices and habits}. Building visualizations from the bottom up generally requires more effort and does not support reliance on reusable templates or other easily chosen solutions. Mendez et al. \cite{Mendez_Hinrichs_Nacenta_2017} confirmed this idea in an empirical study with visualization designers, finding that bottom-up visualization construction requires more thoughtful engagement with the problem and more creativity in the output. However, generating visualizations this way requires both conceptual and technical expertise---a requirement that not all practitioners have. There is not an simple answer to this issue, and we do not claim to provide one here. However, we believe that approaching this topic through the lens of fixation adds a valuable perspective that has not been addressed in the extant literature.

Another topic for future investigation is how to provide inspiration to designers without encouraging fixation. Some participants have developed their own strategies of looking only at visualization examples that are not too similar to the context in which they are currently working. Here there may be an opportunity to develop an inspiration tool or catalog based on certain characteristics of the design situation. For instance, designers could identify characteristics of their design problem, such as data types, complexity, user tasks, and so on, and the tool could present only existing solutions that do not match the characteristics too closely. Perhaps a strategy analogous to Cheng et al. \cite{cheng_new_2014}, where they presented only partial photographs to designers, could be employed to mitigate fixation. Partial visualizations, or visualizations with only partial datasets, could be presented as a strategy to provide inspiration but still require effortful engagement with the ideas. Some of our participants noted looking at art or nature for inspiration. Similarly, an inspiration tool could suggest visual metaphors from nature or paintings with certain color palettes or shapes based on input from the designer. Experimenting with creative exercises like sketching, playing with visual variables, and consciously trying to create bad designs also helped participants overcome fixation. Tools that promote serendipitous mashups of visual variables or provide intentionally `bad' outcomes could aid in this regard. Strategies from the literature on creativity support tools, such as promoting failure in design \cite{kim2015designing}, are likely fruitful for exploration. Although our work does not provide such solutions, our findings surface these topics as potential areas of future research.

\section{Summary}
Our findings indicate that design fixation is prevalent among data visualization practitioners and significantly impacts creativity. We provide initial contributions to the topic of design fixation, including introducing a new theoretical construct to the visualization literature, and partially characterizing it based on an empirical qualitative investigation with professional data visualization practitioners. This work may be valuable for practitioners, helping them develop awareness of fixation in their own practice and collecting some solutions for overcoming it. Our work opens new questions for the research community, including how to support creativity in efficient, flexible ways without encouraging fixation; how to provide useful examples for inspiration without promoting adherence to them; how to encourage designers to `step back' and see the big picture of what they're working on; and how to support perspective shifts to overcome fixation and encourage creativity.


\bibliographystyle{abbrv-doi}

\bibliography{template}
\end{document}